\documentclass[aps,prb,superscriptaddress,nofootinbib,reprint,floatfix]{revtex4-2}

\usepackage{amsmath,amssymb,bm}
\usepackage{graphicx}
\usepackage{xcolor}
\usepackage{placeins}
\usepackage{balance}
\usepackage[colorlinks=true,linkcolor=blue,citecolor=blue,urlcolor=blue]{hyperref}

\hypersetup{
  pdftitle={Interior skin focusing and directional mirror transfer in a graded non-Hermitian Krawtchouk network},
  pdfauthor={Y. S. Liu and X. Z. Zhang}
}

\newcommand{\ii}{\mathrm{i}}
\newcommand{\ee}{\mathrm{e}}
\newcommand{\diag}{\mathrm{diag}}
\newcommand{\OBC}{\mathrm{OBC}}
\newcommand{\ket}[1]{|#1\rangle}
\newcommand{\bra}[1]{\langle #1|}

\begin{document}

\title{Interior skin focusing and directional mirror transfer in a graded non-Hermitian Krawtchouk network}

\author{Y. S. Liu}
\affiliation{College of Physics and Materials Science, Tianjin Normal University, Tianjin 300387, China}

\author{X. Z. Zhang}
\email{zhangxz@tjnu.edu.cn}
\affiliation{College of Physics and Materials Science, Tianjin Normal University, Tianjin 300387, China}
\affiliation{Interdisciplinary Center, Tianjin Normal University, Tianjin 300387, China}

\begin{abstract}
Spatially graded nonreciprocity can move skin weight away from a boundary, but it does not generally preserve a real commensurate spectrum or analytically controlled dynamics. We study a finite open Krawtchouk network with oppositely graded directed hoppings. For a broad intermediate range of asymmetries, their local imaginary gauge field changes sign in the bulk, and its accumulated coordinate generates a positive diagonal similarity map whose normalized squared entries form a biased-binomial envelope. The same map converts the open chain into the spin-rotation generator \(2gJ_x\), so the focus and equally spaced spectrum follow from one grading. In this focusing regime, exact right, left, and biorthogonal eigenvectors show that the envelope width and participation number both scale as \(\sqrt N\), identifying a subextensive interior focus. In the physical node basis, spin rotation produces perfect mirror inversion with direction-selective amplification and attenuation whose gains are mutually inverse. The directional Green functions share their poles, while their residues differ by the same similarity ratio. Closing the chain exposes a gauge-invariant imaginary flux, and either exact one-way limit yields an exceptional point of order \(N\). Onsite disorder preserves the directional resolvent ratio, whereas independent hopping disorder breaks the clean analytic Krawtchouk map and degrades commensurability and transfer. The model therefore provides an exactly solvable finite-network framework linking localization geometry and eigenvector nonorthogonality to commensurate spectra and node-resolved directional response.
\end{abstract}

\maketitle

\section{Introduction}
\label{sec:introduction}

Nonreciprocal hopping acts as an imaginary gauge field and makes spectra, eigenvectors, and responses sensitive to boundary conditions \cite{HatanoNelson1996,HatanoNelson1997,YaoWang2018,Yokomizo2019,Lee2016,liuGradedHoppingScreens2026}. Under open boundary conditions, a uniform gauge field produces boundary skin accumulation, while right and left eigenvectors acquire opposite spatial biases. Non-Bloch spectra describe this boundary-dependent spectrum, while biorthogonal eigenvectors and their nonorthogonality determine propagation and driven response \cite{Kunst2018,Okuma2020,Borgnia2020,Yang2020}. Non-Hermitian symmetry classifications and topology organize these spectral and state structures \cite{Gong2018,Kawabata2019,GhatakDas2019,Ashida2020,Bergholtz2021}. Their boundary-dependent consequences have been examined in electrical, photonic, mechanical, acoustic, and quantum-dynamical platforms \cite{Helbig2020,Zou2021,Zhu2023,Weidemann2020,Ghatak2020,Xiao2020,Zhang2021Acoustic}.

Spatially nonuniform couplings provide additional control over skin profiles. Recent constructions confine skin weight within a finite region of a semi-infinite synthetic lattice \cite{Deng2026Confined} or imprint prescribed profiles through spatially patterned imaginary gauge phases \cite{Dong2026GaugeImprint}. The unresolved issue is therefore not whether an interior profile can occur, but whether it can coexist with a real commensurate spectrum, exact finite-time transfer, and analytic frequency-domain response in one finite network.

We address this problem using oppositely graded directed hoppings \(u_j=\alpha_1j\) and \(v_j=\alpha_2(N-j)\). In the finite-chain focusing regime defined below, their local imaginary gauge field \(\gamma_j=\tfrac12\ln(v_j/u_j)\) changes sign at an interior bond. The accumulated coordinate \(X_j=\sum_{q<j}\gamma_q\) generates a positive diagonal map \(S=\diag(\ee^{X_j})\), whose squared entries, after normalization, form a biased-binomial envelope. The same map converts the open network to the Hermitian Krawtchouk mirror chain \(H_e=2gJ_x\). This Hermitian partner has an equally spaced spectrum and produces exact mirror inversion. The interior focus and the commensurate spectrum therefore follow from a single graded construction.

The exact solution separates the right, left, and biorthogonal eigenvectors and, within the focusing regime, gives a subextensive interior focus whose width grows as \(\sqrt N\). Its dynamical counterpart is perfect mirror inversion. Ordinary node intensity is amplified in one direction and attenuated in reverse. In the frequency domain, the same similarity ratio leaves the pole positions unchanged but rescales their residues. Closing the chain exposes an imaginary-flux obstruction, while the two exact one-way limits form exceptional points of order \(N\). Weak-disorder perturbation theory and finite-size numerics further show that onsite disorder preserves the directional resolvent ratio, whereas independent hopping disorder breaks the clean analytic Krawtchouk map. The diagonal map itself is not specific to this model. The Krawtchouk grading is distinguished by simultaneous closed forms for the similarity envelope, commensurate spectrum, mirror map, and directional Green-function response.

The remainder of this paper is organized as follows. Section~\ref{sec:model} defines the open network and its graded imaginary gauge field. Section~\ref{sec:exact} derives the similarity map, exact spectrum, eigenvectors, focusing laws, and pseudo-Hermiticity. Section~\ref{sec:dynamics} obtains the full propagator, directional mirror transfer, and Green-function response. Section~\ref{sec:robustness} examines disorder, loop closure, and the one-way exceptional-point limits. Section~\ref{sec:discussion} discusses the scope and the bounded-coupling tradeoff. Appendices~\ref{app:response}-\ref{app:loop_ep} provide the collective-ladder construction, detailed derivations, and numerical procedures.

\section{Model and graded nonreciprocity}
\label{sec:model}

We use the orthonormal physical node basis \(\{\ket{j}\}_{j=1}^{N}\), where \(N\ge2\) is the number of nodes and \(j\) labels a locally addressable node of a linear network. Open boundary conditions mean that bonds connect only \(j\) and \(j+1\) for \(j=1,\ldots,N-1\). The Hamiltonian is
\begin{equation}
H_{\OBC}=\sum_{j=1}^{N-1}\left(u_j\ket{j}\bra{j+1}+v_j\ket{j+1}\bra{j}\right),
\label{eq:Hobc}
\end{equation}
with real directed hoppings
\begin{equation}
u_j=\alpha_1j,\qquad v_j=\alpha_2(N-j).
\label{eq:hoppings}
\end{equation}
The coefficient \(u_j\) transfers amplitude from \(j+1\) to \(j\), while \(v_j\) transfers amplitude from \(j\) to \(j+1\). Unless stated otherwise, we take \(\alpha_1>0\) and \(\alpha_2>0\) and define
\begin{equation}
\begin{gathered}
M=N-1,\qquad J=\frac{M}{2},\qquad
g=\sqrt{\alpha_1\alpha_2},\\
r=\frac{\alpha_2}{\alpha_1},\qquad
p=\frac{r}{1+r}.
\end{gathered}
\label{eq:parameters}
\end{equation}

On bond \(j\), the local imaginary gauge field is
\begin{equation}
\gamma_j=\frac12\ln\frac{v_j}{u_j}
=\frac12\ln\left[\frac{r(N-j)}{j}\right].
\label{eq:gamma}
\end{equation}
We define its accumulated coordinate by
\begin{equation}
X_1=0,\qquad X_j=\sum_{q=1}^{j-1}\gamma_q
\quad (j=2,\ldots,N).
\label{eq:Xj}
\end{equation}
The field vanishes at the continuous bond coordinate \(j_\gamma=Np=Nr/(1+r)\). For a finite chain, this zero lies strictly between the first and last bonds when
\begin{equation}
\frac{1}{N-1}<r<N-1.
\label{eq:focusing_regime}
\end{equation}
In this focusing regime, \(X_j\) increases while \(\gamma_j>0\) and decreases after \(\gamma_j<0\), producing an interior maximum rather than monotonic boundary accumulation. At the limiting values in Eq.~\eqref{eq:focusing_regime}, the maximum first reaches an edge bond; beyond them it lies at, or directly adjoins, an endpoint. Varying \(r\) therefore connects interior focusing continuously to boundary accumulation and, ultimately, to the exact one-way limits.

Equation~\eqref{eq:Hobc} defines the effective response matrix of a programmable linear network. At a complex spectral parameter \(z\), local drives \(\bm f\) and measured node amplitudes \(\bm\psi\) obey
\begin{equation}
(zI-H_{\OBC})\bm\psi=\bm f,
\qquad
\bm\psi=G(z)\bm f,
\label{eq:response_equation}
\end{equation}
where \(I\) is the \(N\times N\) identity and \(G(z)=(zI-H_{\OBC})^{-1}\). A drive of amplitude $f_0$ applied only at node $j$ gives $\psi_{\ell}=f_0 G_{\ell j}(z)$. Local drives and readout ports therefore fix the physical node basis, so the nonunitary transformation changes measurable amplitudes and driven response. The factors \(j\) and \(N-j\) also arise in a two-mode collective occupation ladder with fixed total occupation \(M\), providing a combinatorial origin for the Krawtchouk grading \cite{KarlinMcGregor1965,DiaconisGriffiths2012,Griffiths2016}. Appendix~\ref{app:response} distinguishes the normalized and unnormalized Dicke bases.

For \(\alpha_1\alpha_2>0\), a positive diagonal similarity map exists and the spectrum is real, up to an overall sign when both couplings are negative. For \(\alpha_1\alpha_2<0\), the corresponding ladder lies on the imaginary axis. The singular cases \(\alpha_1\alpha_2=0\) are the one-way exceptional-point limits treated in Sec.~\ref{sec:oneway} and Appendix~\ref{app:loop_ep}.

\section{Exact solution and interior focusing}
\label{sec:exact}

\subsection{Similarity map and commensurate spectrum}
\label{sec:similarity}

Define the positive diagonal similarity map
\begin{equation}
S=\diag(s_1,s_2,\ldots,s_N),\qquad s_j>0,\qquad s_1=1.
\label{eq:S}
\end{equation}
Equal transformed hoppings on every open bond require
\begin{equation}
\frac{s_{j+1}}{s_j}=\sqrt{\frac{v_j}{u_j}}=\ee^{\gamma_j}.
\label{eq:srecurrence}
\end{equation}
Equations~\eqref{eq:Xj} and \eqref{eq:srecurrence} give \(s_j=\ee^{X_j}\). Repeated multiplication yields
\begin{equation}
s_j^2=\prod_{q=1}^{j-1}\frac{v_q}{u_q}
=r^{j-1}\binom{N-1}{j-1}.
\label{eq:sprofile}
\end{equation}
The complete product and both transformed hopping directions are shown in Appendix~\ref{app:similarity}.

Define the symmetric transformed hopping by \(\tau_j\equiv\sqrt{u_jv_j}=g\sqrt{j(N-j)}\). Then
\begin{align}
H_e&=S^{-1}H_{\OBC}S\nonumber\\
&=g\sum_{j=1}^{N-1}\sqrt{j(N-j)}
\left(\ket{j}\bra{j+1}+\mathrm{H.c.}\right)\nonumber\\
&=2gJ_x.
\label{eq:He}
\end{align}
Here \(J_x\) is the \(x\) component of a spin \(J=(N-1)/2\), and the node-spin correspondence is \(\ket{j}=\ket{J,m_j}\) with \(m_j=J+1-j\). The exact spectrum is the real commensurate ladder
\begin{equation}
E_\mu=2g\mu,\qquad \mu=-J,-J+1,\ldots,J.
\label{eq:spectrum}
\end{equation}
The construction works bond by bond because an open chain has no additional loop-consistency condition.

\subsection{Eigenvectors and subextensive focusing}
\label{sec:focusing}

Let \(\ket{\phi_\mu}\) be a Euclidean-normalized eigenvector of \(H_e\), so that \(\langle\phi_\nu|\phi_\mu\rangle=\delta_{\nu\mu}\). With \(s_1=1\), we define the canonical right and left eigenvectors by
\begin{equation}
\ket{\psi_\mu^R}=S\ket{\phi_\mu},\qquad
\ket{\psi_\mu^L}=S^{-1}\ket{\phi_\mu}.
\label{eq:RL}
\end{equation}
They obey \(\langle\psi_\nu^L|\psi_\mu^R\rangle=\delta_{\nu\mu}\). Their canonical node-resolved densities are
\begin{align}
\rho_\mu^R(j)&=s_j^2|\phi_\mu(j)|^2,\nonumber\\
\rho_\mu^L(j)&=s_j^{-2}|\phi_\mu(j)|^2,\nonumber\\
\rho_\mu^{\rm bio}(j)&=|\phi_\mu(j)|^2.
\label{eq:densities}
\end{align}
The first two densities are not individually normalized in the ordinary Euclidean norm. Completeness of the Hermitian eigenvectors gives
\begin{align}
\mathcal W_R^{\rm can}(j)&=\sum_\mu|\psi_\mu^R(j)|^2=s_j^2,\nonumber\\
\mathcal W_L^{\rm can}(j)&=\sum_\mu|\psi_\mu^L(j)|^2=s_j^{-2},\nonumber\\
\mathcal W_{\rm bio}(j)&=\sum_\mu\psi_\mu^L(j)^*\psi_\mu^R(j)=1.
\label{eq:integratedweights}
\end{align}
These canonical basis-integrated weights are exact in the stated eigenvector convention and serve as structural diagnostics of the similarity representation. They change under independent mode-wise rescaling, whereas the propagator and resolvent are invariant because the reciprocal rescalings of the left and right spectral factors cancel. A rescaling-invariant measure of mode nonorthogonality is the Petermann factor
\begin{align}
K_\mu
&=\frac{\langle\psi_\mu^R|\psi_\mu^R\rangle
\langle\psi_\mu^L|\psi_\mu^L\rangle}
{|\langle\psi_\mu^L|\psi_\mu^R\rangle|^2}\nonumber\\
&=\left(\sum_j s_j^2|\phi_\mu(j)|^2\right)
\left(\sum_j s_j^{-2}|\phi_\mu(j)|^2\right).
\label{eq:Petermann}
\end{align}
Appendix~\ref{app:similarity} gives the proof and the mode-wise rescaling freedom.

The normalized right-state similarity envelope is
\begin{align}
\varpi_j&=\frac{s_j^2}{\sum_{\ell=1}^{N}s_\ell^2}\nonumber\\
&=\binom{N-1}{j-1}p^{j-1}(1-p)^{N-j}.
\label{eq:varpi}
\end{align}
It is a binomial distribution in \(j-1\), with exact center and variance
\begin{equation}
j_c=1+(N-1)p,\qquad
\sigma_j^2=(N-1)p(1-p).
\label{eq:centerwidth}
\end{equation}
At fixed \(0<p<1\), the relative width scales as \(\sigma_j/N\propto N^{-1/2}\). A large-\(N\) Gaussian approximation further gives the envelope inverse participation ratio
\begin{equation}
\mathcal I_S\equiv\sum_j\varpi_j^2
\sim\frac{1}{2\sqrt{\pi(N-1)p(1-p)}},
\label{eq:IPR}
\end{equation}
and therefore the participation number \(\mathcal P_S=\mathcal I_S^{-1}\propto\sqrt N\). Within the finite-chain interval in Eq.~\eqref{eq:focusing_regime}, we refer to this scaling as subextensive interior skin focusing. Its \(O(\sqrt N)\) width is narrower than an extended \(O(N)\) profile and broader than a conventional \(O(1)\)-width exponential boundary skin mode. The Gaussian and large-deviation derivations, including their range of validity, are given in Appendix~\ref{app:scaling}. Figure~\ref{fig:graded_focus} summarizes the graded gauge field, normalized similarity envelope, canonical basis-integrated weights, and finite-size scaling.

\begin{figure*}[!t]
\includegraphics[width=0.96\textwidth]{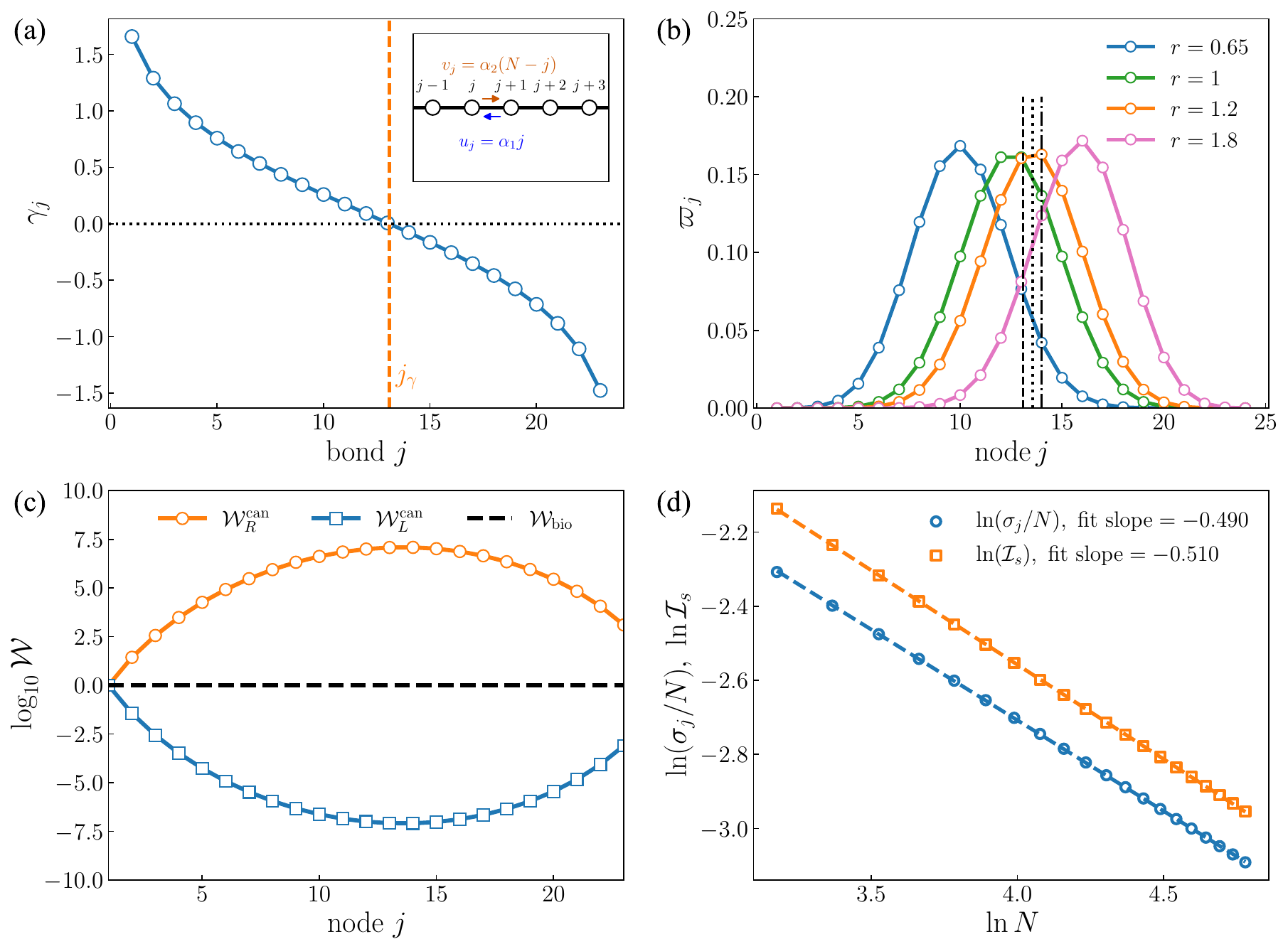}
\caption{Graded imaginary gauge field and the resulting similarity structure. (a) Directed hoppings and \(\gamma_j\) for \(N=24\) and \(r=1.2\), which lies in the finite-chain focusing regime; \(j_\gamma=Np=13.09\) marks the sign change. The inset schematically depicts the Hamiltonian in Eq.~\eqref{eq:Hobc} for a nonreciprocal graded hopping chain, with orange and blue arrows denoting the rightward hopping amplitude \(v_j = \alpha_2(N-j)\) and the leftward hopping amplitude \(u_j = \alpha_1 j\), respectively. (b) Normalized envelopes \(\varpi_j\). For \(r=1.2\), the line styles mark \(j_\gamma=13.09\), \(j_c=13.55\), and the discrete maximizer \(j_{\max}=14\). (c) Canonical basis-integrated weights \(\mathcal W_R^{\rm can}=s_j^2\), \(\mathcal W_L^{\rm can}=s_j^{-2}\), and \(\mathcal W_{\rm bio}=1\), plotted as \(\log_{10}\mathcal W\). The convention is \(s_1=1\) with Euclidean-normalized \(\ket{\phi_\mu}\). (d) Finite-size scaling of \(\sigma_j/N\) and \(\mathcal I_S\) at \(r=1.2\) on logarithmic axes. Independent fits over the displayed sizes give slopes \(-0.490\) and \(-0.510\), consistent with \(-1/2\).}
\label{fig:graded_focus}
\end{figure*}

\begin{figure*}[!t]
\includegraphics[width=0.96\textwidth]{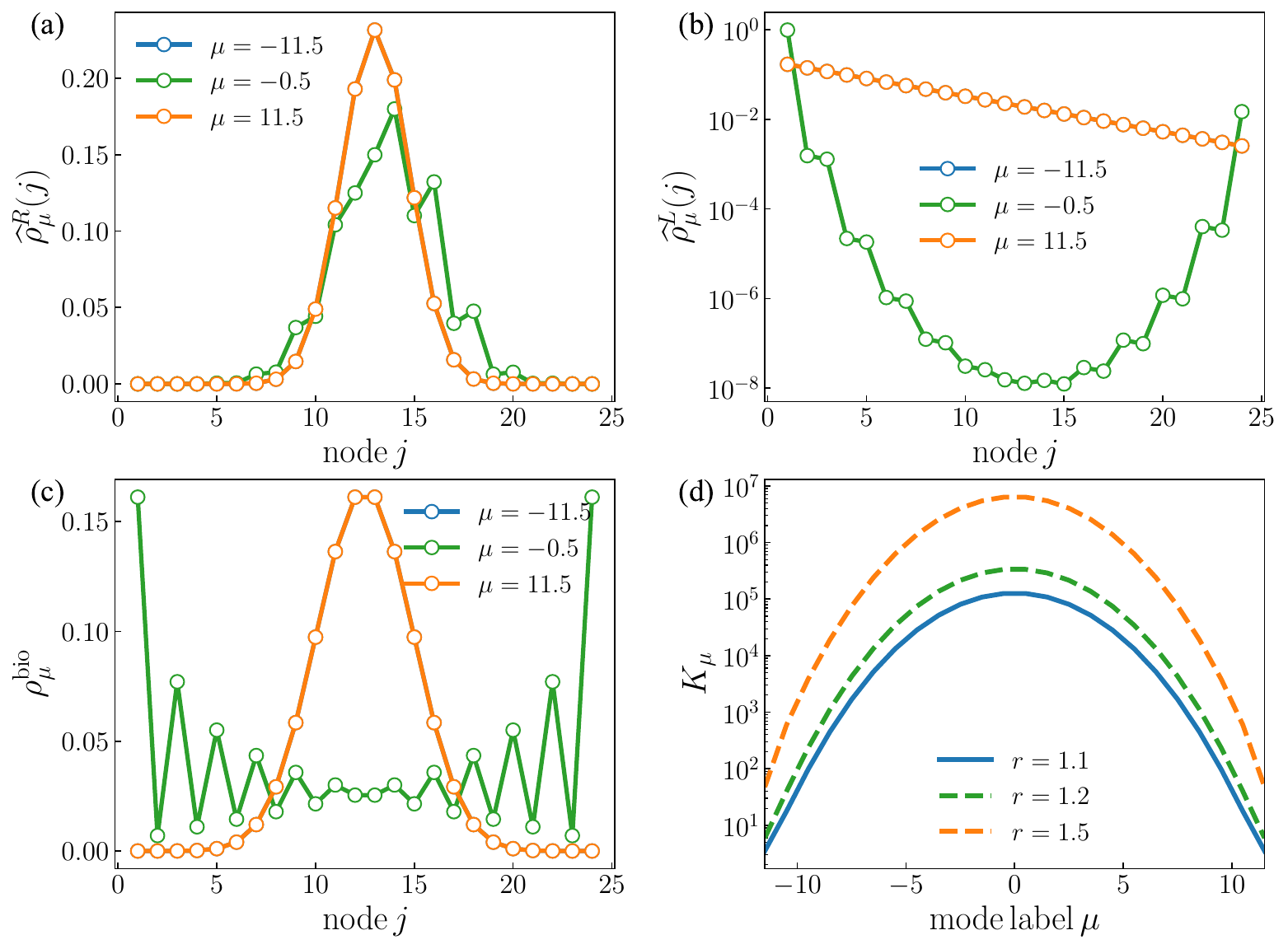}
\caption{Mode-resolved eigenvector structure for \(N=24\). (a,b) Euclidean-normalized right and left densities at \(r=1.2\) for three modes; panel (b) uses a logarithmic y axis. (c) Biorthogonal density \(\rho_\mu^{\rm bio}(j)=|\phi_\mu(j)|^2\). (d) Rescaling-invariant Petermann factor \(K_\mu\) for the displayed values of \(r\), shown on a logarithmic y axis. Colors identify the modes in panels (a)-(c), and line styles identify \(r\) in panel (d). In panels (a)-(c), the orange line overlaps with the blue line.}
\label{fig:mode_resolved}
\end{figure*}

For mode-shape comparisons, we use the Euclidean-normalized densities
\begin{equation}
\widehat\rho_\mu^{R,L}(j)=
\frac{|\psi_\mu^{R,L}(j)|^2}
{\|\psi_\mu^{R,L}\|_2^2}.
\label{eq:normalized_mode_densities}
\end{equation}
Figure~\ref{fig:mode_resolved} shows how the common similarity envelope is modulated by the Krawtchouk nodal structure. The reciprocal similarity factors bias the Euclidean-normalized right and left profiles toward different spatial regions, while the detailed nodal structure remains mode dependent. The biorthogonal density retains the corresponding Hermitian Krawtchouk profile. The growth of \(K_\mu\) away from \(r=1\) quantifies the accompanying eigenvector nonorthogonality.

The physical node basis is fixed by local ports or fields, so the nonunitary similarity transformation changes observable node amplitudes even though it preserves the clean OBC eigenvalues.

\subsection{Pseudo-Hermiticity}
\label{sec:pseudo}

The positive diagonal pseudo-Hermitian metric is \cite{BenderBoettcher1998,Mostafazadeh2002}
\begin{equation}
\eta=S^{-2}=\diag(\ee^{-2X_1},\ldots,\ee^{-2X_N}).
\label{eq:eta}
\end{equation}
Because \(H_e=H_e^\dagger\), the open chain satisfies
\begin{equation}
H_{\OBC}^\dagger\eta=\eta H_{\OBC}.
\label{eq:pseudo}
\end{equation}
For \(U(t)=\exp(-\ii H_{\OBC}t)\), this identity implies
\begin{equation}
U(t)^\dagger\eta U(t)=\eta.
\label{eq:pseudounitary}
\end{equation}
Thus \(\langle\psi(t)|\eta|\psi(t)\rangle\) is conserved, whereas the ordinary node intensity \(\sum_j|\psi_j(t)|^2\) generally is not. This distinction is physical when local ports or local fields define the node basis: the nonunitary map changes time-domain amplitudes and driven Green-function residues. The clean OBC eigenvalues are Hermitian-equivalent, while eigenvector nonorthogonality, node-basis response, directional dynamics, and loop obstruction retain non-Hermitian content. Detailed proofs of Eqs.~\eqref{eq:pseudo} and \eqref{eq:pseudounitary} are in Appendix~\ref{app:similarity}.

\section{Directional dynamics and Green-function response}
\label{sec:dynamics}

Both the propagator and resolvent inherit the prefactor \(s_\ell/s_j\), linking directional transfer in time to residue asymmetry in frequency.

\subsection{Full propagator and mirror inversion}
\label{sec:mirror}

The Hermitian Krawtchouk chain is an engineered mirror-transfer network \cite{Christandl2004,Albanese2004,Christandl2005,VinetZhedanov2012,VinetZhedanov2012Para,VinetZhedanov2012APST,JafarovVanDerJeugt2010,Kay2010,Longhi2014}. Non-Hermitian perfect transfer has also been studied in parity-time-symmetric networks \cite{ZhangJinSong2012}. In the present chain, the full propagator is
\begin{align}
A_{\ell j}(t)&\equiv\langle\ell|U(t)|j\rangle\nonumber\\
&=\frac{s_\ell}{s_j}\langle J,m_\ell|\ee^{-\ii2gJ_xt}|J,m_j\rangle\nonumber\\
&=\frac{s_\ell}{s_j}\,\ii^{m_\ell-m_j}
d^J_{m_\ell m_j}(2gt),
\label{eq:wignerd}
\end{align}
where \(d^J_{m'm}(\beta)=\langle J,m'|\exp(-\ii\beta J_y)|J,m\rangle\) is the standard real Wigner small-\(d\) matrix. Appendix~\ref{app:propagator_resolvent}, Sec.~1 fixes the phase convention, gives the finite factorial sum, and compares Eq.~\eqref{eq:wignerd} with direct matrix exponentiation.

\begin{figure*}[!t]
\includegraphics[width=0.96\textwidth]{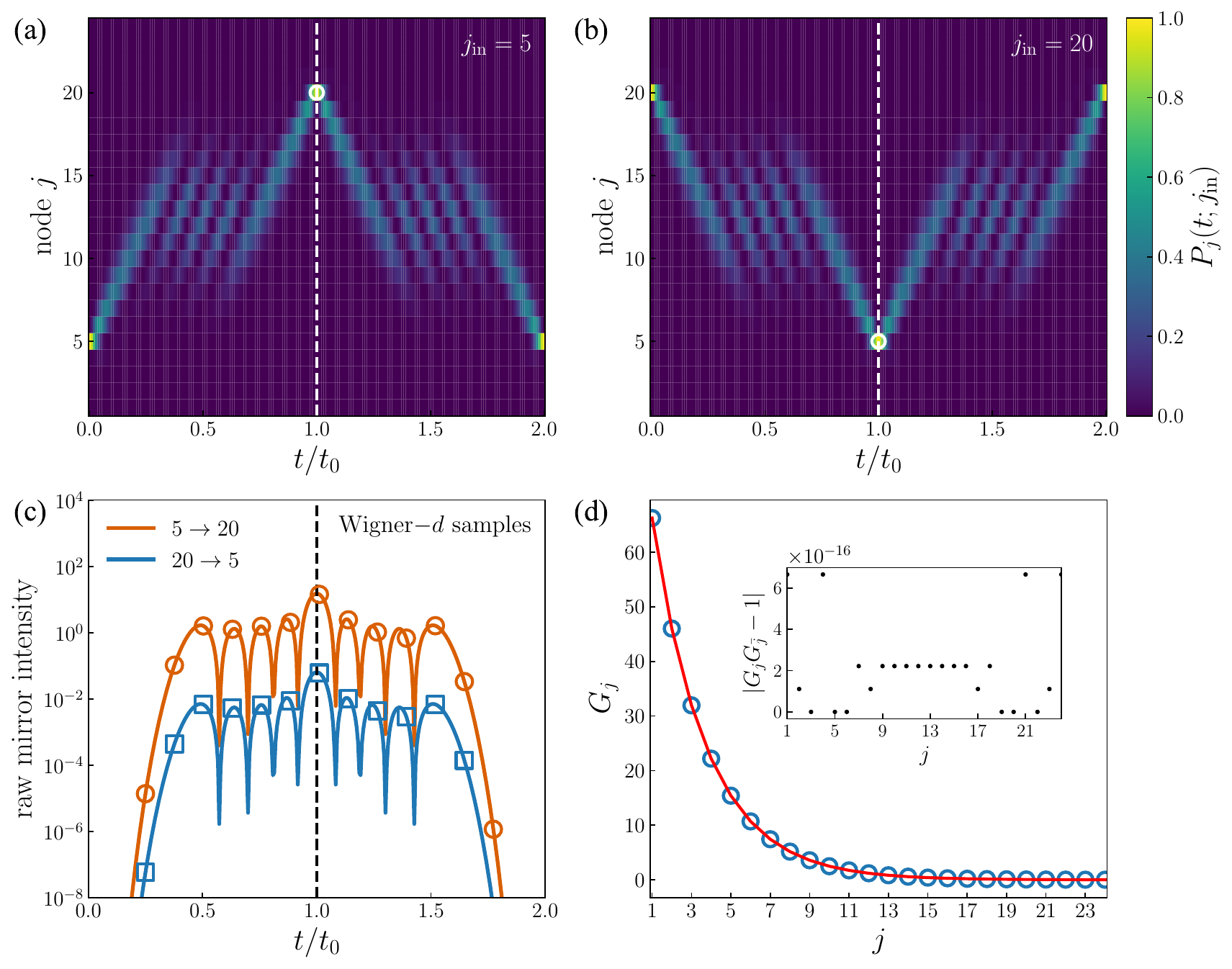}
\caption{Direction-selective mirror dynamics in the same \(N=24\), \(r=1.2\) open network. (a) Node distribution \(P_j(t;5)\). (b) Reverse distribution \(P_j(t;20)\). The two heat maps are independently normalized according to Eq.~\eqref{eq:Pnode}; they show routing but not gain. Gain information is retained in panel (c), which shows the raw mirror intensities from direct matrix exponentiation, analytic Wigner-\(d\) samples, and filled markers at \(t=t_0\). (d) Exact mirror gain \(G_j=r^{N+1-2j}\) and direct numerical values. The inset shows the reciprocal-product residual \( |G_jG_{\bar j}-1| \).}
\label{fig:mirror}
\end{figure*}

At the mirror time
\begin{equation}
t_0=\frac{\pi}{2g},\qquad \bar j=N+1-j,
\label{eq:t0}
\end{equation}
the spin rotation maps every node to its mirror up to the site-independent phase \(\chi_N=(-\ii)^{N-1}\). Therefore
\begin{equation}
A_{\bar j,j}(t_0)=\chi_N\frac{s_{\bar j}}{s_j},\qquad
G_j\equiv|A_{\bar j,j}(t_0)|^2=r^{N+1-2j}.
\label{eq:mirrorgain}
\end{equation}
The reverse gain is \(G_{\bar j}=G_j^{-1}\), so every mirror pair obeys
\begin{equation}
G_jG_{\bar j}=1.
\label{eq:gainproduct}
\end{equation}
The normalized distribution
\begin{equation}
P_\ell(t;j)=\frac{|A_{\ell j}(t)|^2}{\sum_n|A_{nj}(t)|^2}
\label{eq:Pnode}
\end{equation}
isolates routing but removes the raw gain. Figure~\ref{fig:mirror} therefore plots the normalized distributions together with the unnormalized mirror intensities for launches from a mirror pair in one fixed network.

\begin{figure*}[!t]
\includegraphics[width=0.96\textwidth]{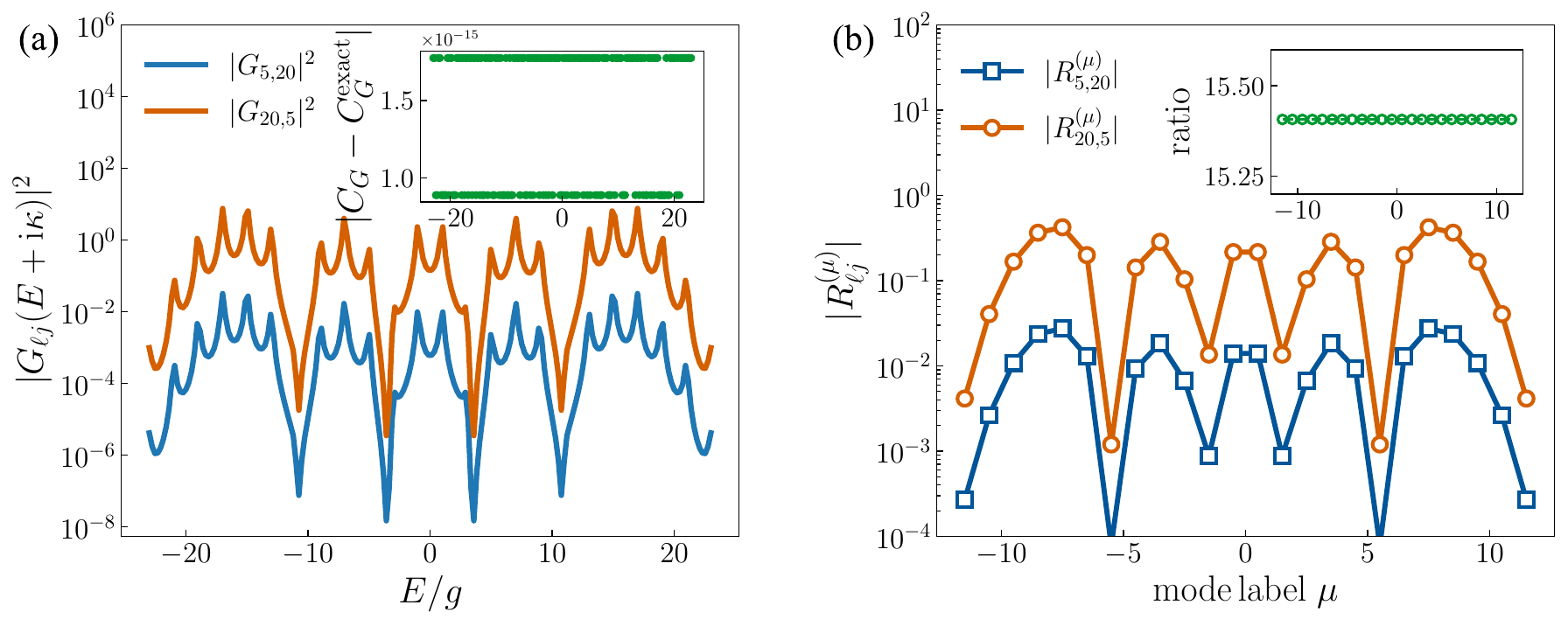}
\caption{Clean Green-function response for \(N=24\), \(r=1.2\), \(j=5\), \(\bar j=20\), and \(\kappa=0.12g\). (a) Directional intensities \( |G_{20,5}(E+\ii\kappa)|^2 \) and \( |G_{5,20}(E+\ii\kappa)|^2 \) on a logarithmic scale. Their resonances occur at the same pole positions. The inset shows the residual \(C_G(E)-C_G^{\rm exact}\), where \(C_G^{\rm exact}=2(N+1-2j)\ln r\). (b) Absolute residues in the two directions on a single logarithmic y axis. The inset shows their mode-independent ratio \( |R_{20,5}^{(\mu)}|/|R_{5,20}^{(\mu)}|=r^{N+1-2j} \).}
\label{fig:green}
\end{figure*}

The frequency-domain counterpart is summarized in Fig.~\ref{fig:green} and derived below.

\subsection{Green-function poles and residues}
\label{sec:green}

Let \(z=E+\ii\kappa\), where \(E\) is a real drive frequency and \(\kappa>0\) is a broadening. The resolvent introduced in Eq.~\eqref{eq:response_equation} satisfies
\begin{equation}
G(z)=(zI-H_{\OBC})^{-1}
=S(zI-H_e)^{-1}S^{-1}.
\label{eq:Green}
\end{equation}
Since the Hermitian resolvent is symmetric in the node basis,
\begin{equation}
\frac{G_{\ell j}(z)}{G_{j\ell}(z)}
=\left(\frac{s_\ell}{s_j}\right)^2.
\label{eq:Greenratio}
\end{equation}
The corresponding logarithmic intensity contrast is
\begin{equation}
C_G(E)\equiv
\ln\frac{|G_{\ell j}(E+\ii\kappa)|^2}
{|G_{j\ell}(E+\ii\kappa)|^2}
=4\ln\frac{s_\ell}{s_j}.
\label{eq:CG}
\end{equation}
Equation~\eqref{eq:Greenratio} applies when the two matrix elements are nonzero and extends across common isolated zeros by analytic continuation.

The pole positions are direction independent, whereas the residues retain the similarity factor. Explicitly,
\begin{equation}
R_{\ell j}^{(\mu)}
=\frac{s_\ell}{s_j}\phi_\mu(\ell)\phi_\mu(j)^*.
\label{eq:residue}
\end{equation}
Writing \(G^{(e)}(z)=(zI-H_e)^{-1}\) for the Hermitian-partner resolvent, its endpoint element is
\begin{equation}
G^{(e)}_{N1}(z)=
\frac{g^{N-1}(N-1)!}
{\displaystyle\prod_{\mu=-J}^{J}(z-2g\mu)},
\label{eq:endpointGreen}
\end{equation}
which makes the common pole set explicit. Appendix~\ref{app:propagator_resolvent}, Sec.~2 derives the spectral and cofactor forms and specializes the directional ratio to mirror pairs. Figure~\ref{fig:green} displays this separation directly: the two directions share the same resonances, while their residues differ by the exact similarity ratio.

The same positive diagonal map relates the eigenvectors, propagator, and resolvent to their Hermitian counterparts. Consequently, the mirror amplitudes in Eq.~\eqref{eq:mirrorgain} and the Green-function residues in Eq.~\eqref{eq:residue} carry the same directional similarity factor, while the pole positions remain those of the commensurate Krawtchouk ladder.

\section{Robustness and singular limits}
\label{sec:robustness}

\subsection{Disorder robustness}
\label{sec:disorder}

We distinguish onsite and relative hopping disorder and use the previously defined \(\tau_j=g\sqrt{j(N-j)}\). Onsite disorder is
\begin{equation}
V_\epsilon=\sum_{j=1}^{N}\epsilon_j\ket{j}\bra{j},
\qquad \epsilon_j\in[-W_\epsilon g,W_\epsilon g],
\label{eq:onsitedisorder}
\end{equation}
where the variables are independent and uniformly distributed. Relative hopping disorder replaces \(u_j\) and \(v_j\) by \(u_j(1+\xi_j^u)\) and \(v_j(1+\xi_j^v)\), respectively, with independent \(\xi_j^{u,v}\in[-W_h,W_h]\). First-order biorthogonal perturbation theory gives
\begin{align}
\operatorname{Var}(\delta E_\mu)_\epsilon
&=\frac{W_\epsilon^2g^2}{3}
\sum_j|\phi_\mu(j)|^4,\nonumber\\
\operatorname{Var}(\delta E_\mu)_h
&=\frac{2W_h^2}{3}\sum_{j=1}^{N-1}
\tau_j^2|\phi_\mu(j)\phi_\mu(j+1)|^2.
\label{eq:disordervariance}
\end{align}
These formulas assume independent zero-mean disorder, nondegenerate clean levels, and perturbatively small \(W_\epsilon\) or \(W_h\). Their derivation is given in Appendix~\ref{app:disorder}.

\begin{figure*}[!t]
\includegraphics[width=0.96\textwidth]{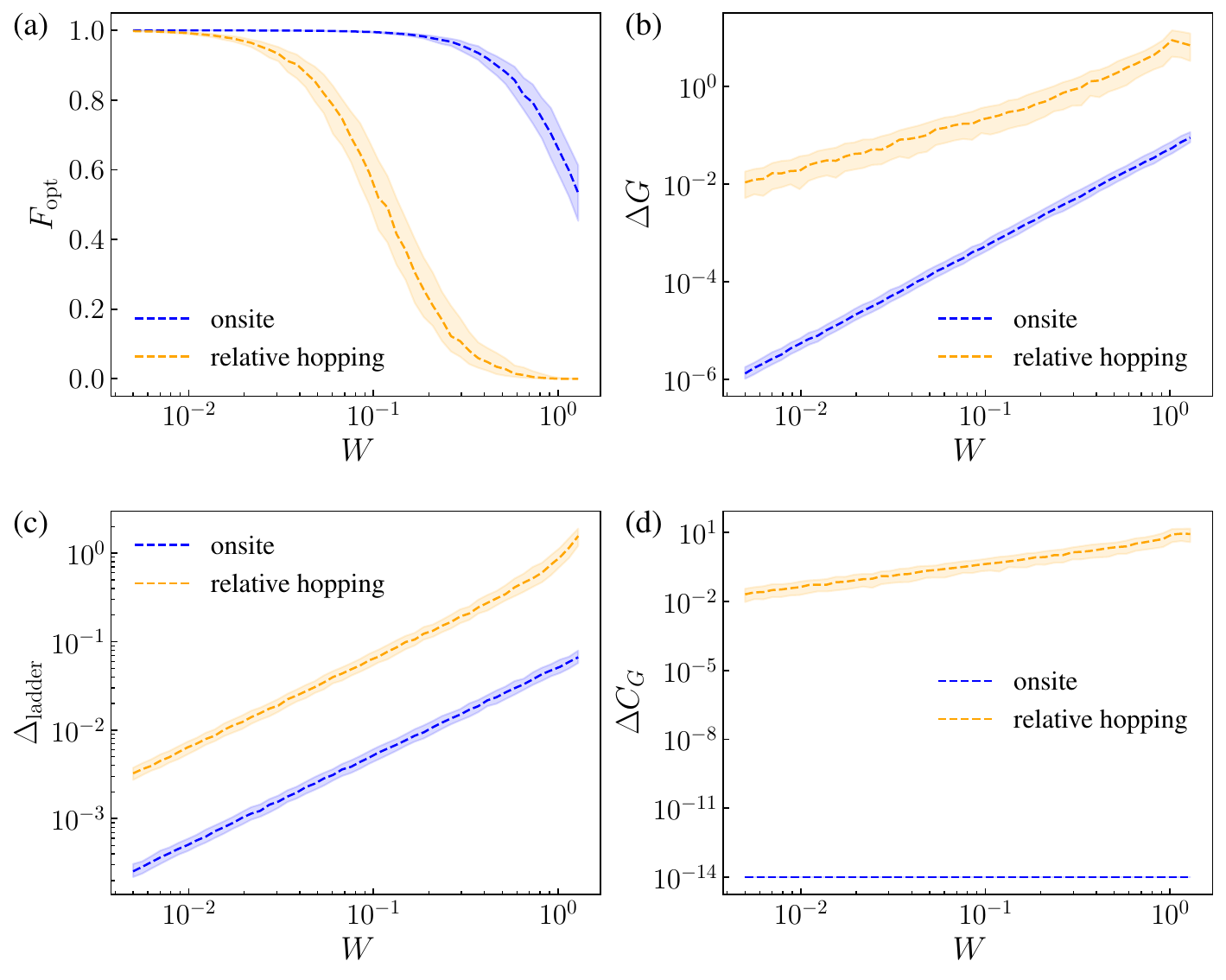}
\caption{Finite-disorder response for \(N=24\), \(r=1.2\), and \(j_0=5\). The common horizontal variable \(W\) denotes \(W_\epsilon\) for onsite disorder and \(W_h\) for relative hopping disorder. (a) Optimized routing fidelity \(F_{\rm opt}\). (b) Log-gain deviation \(\Delta_G\), with \(G_{\rm opt}\) evaluated at the same \(t_*\) that maximizes \(F\). (c) Empirical ladder distortion \(\Delta_{\rm ladder}\). (d) Green-contrast deviation \(\Delta C_G\) at \(E=0\) and \(\kappa=0.12g\); values below \(10^{-14}\) are shown at the numerical floor. Curves are medians and shaded regions are interquartile intervals from 500 fixed-seed realizations. Propagation uses direct matrix exponentiation, and eigenvalues are assigned optimally to the clean ladder before fitting.}
\label{fig:disorder}
\end{figure*}

For a launch at \(j_0\), let
\begin{equation}
F(t)=\frac{|\langle\bar j_0|U(t)|j_0\rangle|^2}
{\sum_\ell|\langle\ell|U(t)|j_0\rangle|^2},
\qquad
t_*=\underset{0.8t_0\le t\le1.2t_0}{\arg\max}\,F(t).
\label{eq:tstar}
\end{equation}
We use \(F_{\rm opt}=F(t_*)\) and the raw optimized gain
\begin{equation}
G_{\rm opt}=|\langle\bar j_0|U(t_*)|j_0\rangle|^2,
\qquad
\Delta_G=|\ln G_{\rm opt}-\ln G_{j_0}|.
\label{eq:Gopt}
\end{equation}
To quantify spectral distortion, disordered eigenvalues are assigned optimally to the clean ladder and fitted to \(\beta_0+\beta_1\mu\), where \(\beta_0\) and \(\beta_1\) are complex fit coefficients. The normalized root-mean-square residual is denoted by \(\Delta_{\rm ladder}\). For the driven response at \(E=0\), we define
\begin{equation}
\Delta C_G=|C_G-C_G^{(0)}|,
\qquad C_G^{(0)}=2(N+1-2j_0)\ln r.
\label{eq:DeltaCG}
\end{equation}
Real onsite disorder commutes with \(S\), so \(S^{-1}(H_{\OBC}+V_\epsilon)S=H_e+V_\epsilon\) remains Hermitian. It can spoil commensurability and mirror inversion, but the clean directional Green ratio remains exact. For realizations with positive disordered bonds, a sample-dependent diagonal map still symmetrizes the OBC chain. However, independent disorder in the two hopping directions breaks the clean analytic Krawtchouk map, and the transformed hoppings no longer have the form \(g\sqrt{j(N-j)}\). The exact commensurate ladder and mirror inversion are therefore lost. Figure~\ref{fig:disorder} reports finite-\(N\) numerical evidence and does not imply asymptotic disorder stability. Near \(W_h=1\), individual bonds can approach zero, and \(\Delta_{\rm ladder}\) is only an empirical spectral-distortion measure.

\begin{figure*}[!t]
\includegraphics[width=0.96\textwidth]{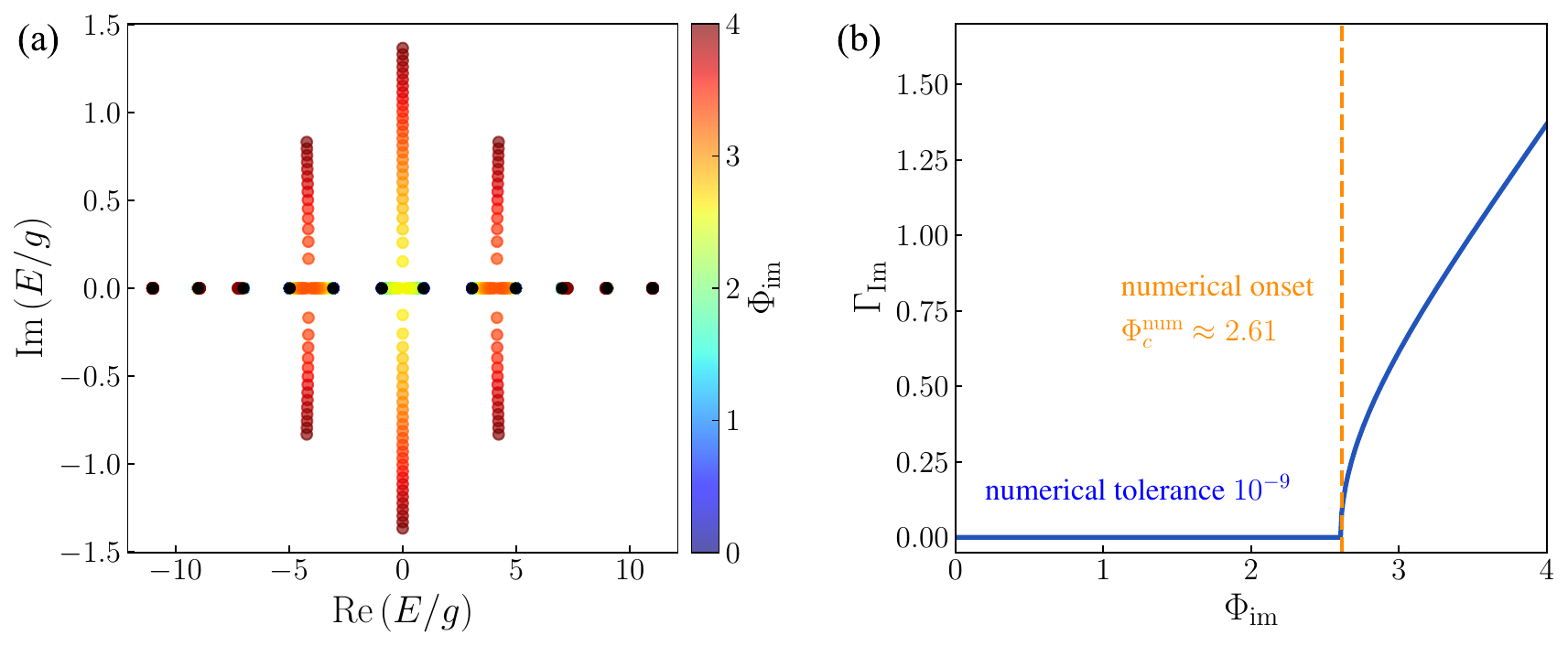}
\caption{Spectral consequence of loop imaginary flux for \(N=12\) and \(\tau_b=0.20g\). (a) Continuously matched complex eigenvalue trajectories as \(\Phi_{\rm im}\) increases from 0 to 4; color encodes the flux and black dots mark \(\Phi_{\rm im}=0\). (b) \(\Gamma_{\rm Im}=g^{-1}\max_n|\operatorname{Im}E_n|\). The dashed line marks the first sampled flux for which \(\Gamma_{\rm Im}>10^{-9}\). This numerical onset depends on \(N\), \(\tau_b\), the flux grid, and the chosen tolerance. The nonzero flux is the exact gauge obstruction, whereas the onset of finite imaginary parts is a finite-size spectral property.}
\label{fig:loop_flux}
\end{figure*}

The loop-spectrum diagnostic is shown in Fig.~\ref{fig:loop_flux} and discussed next. Full disorder distributions and sample-number convergence are given in Fig.~\ref{fig:appendix_disorder} and Appendix~\ref{app:disorder}.

\subsection{Closed-loop obstruction}
\label{sec:loop}

We close the chain with
\begin{equation}
\begin{aligned}
H_b={}&w_{1\leftarrow N}\ket{1}\bra{N}
+w_{N\leftarrow1}\ket{N}\bra{1},\\
H_{\rm ring}={}&H_{\OBC}+H_b,
\end{aligned}
\label{eq:Hb}
\end{equation}
where both closing amplitudes are positive. The signed imaginary flux is
\begin{equation}
\Phi_{\rm im}=\sum_{j=1}^{N-1}\gamma_j
+\frac12\ln\frac{w_{1\leftarrow N}}{w_{N\leftarrow1}}.
\label{eq:flux}
\end{equation}
Writing \(\tau_b=\sqrt{w_{1\leftarrow N}w_{N\leftarrow1}}\), the transformed closing amplitudes are \(\tau_b\ee^{\Phi_{\rm im}}\) and \(\tau_b\ee^{-\Phi_{\rm im}}\). Hermiticity requires \(\Phi_{\rm im}=0\). A nonzero \(\Phi_{\rm im}\) is therefore the exact obstruction to extending the open-chain similarity transformation around the loop. This algebraic obstruction does not imply that an arbitrarily small flux immediately produces complex eigenvalues in a finite ring. The onset of spectral complexification depends on the closing strength and system size. Figure~\ref{fig:loop_flux}(a) tracks the continuously matched eigenvalue trajectories as the flux is increased, while Fig.~\ref{fig:loop_flux}(b) shows the corresponding maximum imaginary part. Appendix~\ref{app:loop_ep} gives the loop-consistency proof and trajectory-assignment procedure.

\subsection{One-way exceptional points}
\label{sec:oneway}

The one-way exceptional point is a distinct singular limit that does not require boundary closure. It arises because the open-chain similarity map loses invertibility as one hopping direction vanishes. At \(\alpha_2=0\) with \(\alpha_1\ne0\), the open-chain matrix is strictly upper triangular and obeys
\begin{equation}
H_{\OBC}^{N}=0,\qquad
H_{\OBC}^{N-1}=\alpha_1^{N-1}(N-1)!\ket{1}\bra{N}\ne0.
\label{eq:nilpotent}
\end{equation}
It therefore has nilpotency index \(N\): all eigenvalues coalesce at zero, the geometric multiplicity is one, and the matrix forms a single Jordan block of size \(N\). The case \(\alpha_1=0\), \(\alpha_2\ne0\) is its lower-triangular counterpart. These claims are restricted to the exact one-way limits. Exceptional points and their response implications are reviewed in Refs.~\cite{Heiss2012,MiriAlu2019}, while sensing studies emphasize the distinction between spectral splitting and full input-output response \cite{Wiersig2014,Hodaei2017,Chen2017,LauClerk2018,McDonaldClerk2020,BudichBergholtz2020}.

For \(r>0\), let \(V^{\rm can}=S\Phi\), where the columns of the unitary matrix \(\Phi\) are the vectors \(\ket{\phi_\mu}\). The canonical similarity-conditioning scale is exactly
\begin{equation}
\begin{aligned}
\kappa_S&\equiv\operatorname{cond}_2(S)
=\operatorname{cond}_2(V^{\rm can})\\
&=\exp\left(\max_jX_j-\min_jX_j\right).
\end{aligned}
\label{eq:kappaS}
\end{equation}
It is evaluated in log space and does not rely on an ill-conditioned double-precision eigensolver. As \(r\to0^+\) at fixed \(\alpha_1\), the real ladder \(E_\mu=2\alpha_1\sqrt r\,\mu\) collapses and \(\kappa_S\) diverges. The equality $\mathrm{cond}_2(V^{\rm can}) = \mathrm{cond}_2(S)$ refers to the canonical normalization in Eq.~\eqref{eq:RL}. Arbitrary mode-wise rescaling changes the eigenvector condition number, whereas $\kappa_S$ is fixed by the diagonal similarity map up to an irrelevant overall scale; its divergence tracks the singularity of this fixed map. Figure~\ref{fig:appendix_ep} and Appendix~\ref{app:loop_ep} compare the exact collapse of the spectral width with the divergence of the canonical similarity condition number and give the Jordan proof and high-precision validation.

\section{Discussion and outlook}
\label{sec:discussion}

The grading plays two distinct roles. Within \(1/(N-1)<r<N-1\), the sign reversal of \(\gamma_j\) places the maximum of \(X_j\) in the bulk, fixing mutually inverse right- and left-state envelopes \(s_j^2\) and \(s_j^{-2}\). Outside this interval the maximum moves to an endpoint, continuously connecting interior focusing to boundary accumulation. At the same time, the bond products \(u_jv_j\) reproduce the Krawtchouk \(J_x\) couplings and hence the commensurate ladder. Because the same \(S\) relates the physical and Hermitian frames, mirror amplitudes and Green-function residues carry the same directional factor. Boundary closure and one-way hopping expose different limitations of this construction. A loop makes the accumulated imaginary gauge field globally observable through \(\Phi_{\rm im}\), whereas the one-way limit makes the diagonal map itself singular.

These results concern a finite, nonuniform open network, and no Bloch topological invariant is assigned. The focus is subextensive rather than exponentially localized with an \(O(1)\) width. A positive metric requires \(\alpha_1\alpha_2>0\); opposite signs give an imaginary ladder instead. The order-\(N\) exceptional point occurs only at an exact one-way limit. Beyond weak disorder, the reported trends are finite-size numerical evidence rather than perturbative results.

For fixed $g$, the largest Hermitianized hopping scales as $\tau_{\rm max} \sim gN/2$. Keeping local couplings bounded by taking $g = g_0/N$ at fixed $r$ leaves the envelope unchanged, gives $\tau_{\rm max} = O(g_0)$, and increases the mirror time to $t_0 = \pi N/(2g_0)$. Thus bounded local couplings require a transfer time linear in $N$. More generally, graded similarity maps can coordinate spatial response and spectral commensurability subject to this coupling-time tradeoff.

\begin{acknowledgments}
We acknowledge support from the National Natural Science Foundation of China under Grants Nos. 12675024 and 12275193.
\end{acknowledgments}

\section*{Data availability}

The numerical data and Python scripts supporting the findings of this study are available from the corresponding author upon reasonable request.

\appendix

\section{Collective-ladder origin}
\label{app:response}

The linear factors in Eq.~\eqref{eq:hoppings} arise directly in a collective two-mode ladder. Let \(a\) and \(b\) be bosonic annihilation operators satisfying \([a,a^\dagger]=[b,b^\dagger]=1\), with \([a,b]=[a,b^\dagger]=[a^\dagger,b]=[a^\dagger,b^\dagger]=0\). Fix the total occupation \(a^\dagger a+b^\dagger b=M\). The unnormalized states are
\begin{equation}
\ket{n}_{\rm u}=(a^\dagger)^{M-n}(b^\dagger)^n\ket{0},
\qquad n=0,1,\ldots,M.
\label{eq:unnormalized_dicke}
\end{equation}
Using
\begin{equation}
a(a^\dagger)^q=(a^\dagger)^qa+q(a^\dagger)^{q-1},
\label{eq:boson_identity}
\end{equation}
and \(a\ket0=b\ket0=0\), one obtains
\begin{align}
b^\dagger a\ket{n}_{\rm u}
&=b^\dagger a(a^\dagger)^{M-n}(b^\dagger)^n\ket0\nonumber\\
&=(M-n)b^\dagger(a^\dagger)^{M-n-1}(b^\dagger)^n\ket0\nonumber\\
&=(M-n)\ket{n+1}_{\rm u},
\label{eq:ba_unnormalized}
\end{align}
and
\begin{align}
a^\dagger b\ket{n}_{\rm u}
&=a^\dagger b(a^\dagger)^{M-n}(b^\dagger)^n\ket0\nonumber\\
&=n a^\dagger(a^\dagger)^{M-n}(b^\dagger)^{n-1}\ket0\nonumber\\
&=n\ket{n-1}_{\rm u}.
\label{eq:ab_unnormalized}
\end{align}
Labeling node $j$ by the unnormalized occupation coordinate $n = j - 1$, the corresponding coefficient representation of $H_{\rm coll} = \alpha_1 a^\dagger b + \alpha_2 b^\dagger a$ gives the adjacent transitions
\begin{align}
H_{\rm coll}\ket{n=j}_{\rm u}
&\supset \alpha_1j\ket{n=j-1}_{\rm u},\nonumber\\
H_{\rm coll}\ket{n=j-1}_{\rm u}
&\supset \alpha_2(N-j)\ket{n=j}_{\rm u}.
\label{eq:collective_node_mapping}
\end{align}
The occupation-coordinate representation therefore reproduces the directed coefficients \(u_j=\alpha_1j\) and \(v_j=\alpha_2(N-j)\). This correspondence is at the level of the coefficient representation; the physical node basis in Eq.~\eqref{eq:Hobc} remains orthonormal. The symbol \(\supset\) isolates the indicated adjacent transition from the full action of \(H_{\rm coll}\).

The normalized Dicke states are
\begin{equation}
\ket n=\frac{(a^\dagger)^{M-n}(b^\dagger)^n}
{\sqrt{(M-n)!n!}}\ket0.
\label{eq:normalized_dicke}
\end{equation}
Applying the same commutator identities and retaining the normalization ratios gives
\begin{align}
b^\dagger a\ket n
&=\sqrt{(M-n)(n+1)}\ket{n+1},\nonumber\\
a^\dagger b\ket n
&=\sqrt{n(M-n+1)}\ket{n-1}.
\label{eq:normalized_ladder}
\end{align}
The normalized matrix elements are the standard spin-ladder square roots. The unnormalized representation supplies the linear directed coefficients, while normalization restores the angular-momentum factors. For \(\alpha_1\ne\alpha_2\), the positive diagonal map introduced in Sec.~\ref{sec:similarity} removes the remaining directional imbalance.

\section{Detailed similarity transformation, eigenvectors, and pseudo-Hermiticity}
\label{app:similarity}

For the diagonal map in Eq.~\eqref{eq:S}, the two terms on bond \(j\) transform as
\begin{align}
S^{-1}\left(u_j\ket j\bra{j+1}\right)S
&=u_j\frac{s_{j+1}}{s_j}\ket j\bra{j+1},
\label{eq:bond_transform_u}\\
S^{-1}\left(v_j\ket{j+1}\bra j\right)S
&=v_j\frac{s_j}{s_{j+1}}\ket{j+1}\bra j.
\label{eq:bond_transform_v}
\end{align}
Equating the two transformed amplitudes gives
\begin{equation}
u_j\frac{s_{j+1}}{s_j}=v_j\frac{s_j}{s_{j+1}}.
\label{eq:symmetry_condition}
\end{equation}
Multiplication by \(s_{j+1}/(u_js_j)\) yields
\begin{equation}
\frac{s_{j+1}^2}{s_j^2}
=\frac{v_j}{u_j}=\frac{r(N-j)}{j}.
\label{eq:squared_recurrence}
\end{equation}
With \(s_1=1\), every product step is
\begin{align}
s_j^2
&=\frac{v_1}{u_1}\frac{v_2}{u_2}\cdots
\frac{v_{j-1}}{u_{j-1}}\nonumber\\
&=\prod_{q=1}^{j-1}
\frac{\alpha_2(N-q)}{\alpha_1q}\nonumber\\
&=r^{j-1}
\frac{(N-1)(N-2)\cdots(N-j+1)}
{1\cdot2\cdots(j-1)}\nonumber\\
&=r^{j-1}\frac{(N-1)!}{(j-1)!(N-j)!}\nonumber\\
&=r^{j-1}\binom{N-1}{j-1}.
\label{eq:full_s_product}
\end{align}
Taking the positive square root of Eq.~\eqref{eq:squared_recurrence} also gives \(s_{j+1}/s_j=\ee^{\gamma_j}\), so \(s_j=\ee^{X_j}\).

Substitution into Eq.~\eqref{eq:bond_transform_u} gives
\begin{align}
u_j\frac{s_{j+1}}{s_j}
&=\alpha_1j\sqrt{\frac{\alpha_2(N-j)}{\alpha_1j}}\nonumber\\
&=\sqrt{\alpha_1\alpha_2}\sqrt{j(N-j)}
=\tau_j.
\label{eq:transformed_u}
\end{align}
Similarly,
\begin{align}
v_j\frac{s_j}{s_{j+1}}
&=\alpha_2(N-j)
\sqrt{\frac{\alpha_1j}{\alpha_2(N-j)}}\nonumber\\
&=\sqrt{\alpha_1\alpha_2}\sqrt{j(N-j)}
=\tau_j.
\label{eq:transformed_v}
\end{align}
This proves the symmetric form of \(H_e\). The open chain has \(N-1\) recurrence conditions for \(N\) positive numbers \(s_j\), leaving only the overall scale free. A loop adds one more condition, as shown in Appendix~\ref{app:loop_ep}.

If \(\alpha_1\alpha_2<0\), the ratio \(v_j/u_j\) is negative and no positive diagonal \(S\) satisfies Eq.~\eqref{eq:squared_recurrence}. A complex diagonal map can instead use \(d_{j+1}/d_j=\sqrt{v_j/u_j}\). With a consistent square-root branch, it transforms the chain to \(\pm2\ii g_-J_x\), where \(g_-=\sqrt{|\alpha_1\alpha_2|}\). The spectrum is therefore \(E_\mu=\pm2\ii g_-\mu\), while the positive metric in Eq.~\eqref{eq:eta} is unavailable. This is distinct from the singular one-way cases, for which the diagonal map ceases to be invertible.

The node-spin mapping follows from \(m_j=J+1-j\). Since \(m_{j+1}=m_j-1\),
\begin{align}
\langle J,m_{j+1}|J_x|J,m_j\rangle
&=\frac12\sqrt{(J+m_j)(J-m_j+1)}\nonumber\\
&=\frac12\sqrt{(N-j)j}.
\label{eq:Jx_element}
\end{align}
Comparison with Eqs.~\eqref{eq:transformed_u} and \eqref{eq:transformed_v} proves \(H_e=2gJ_x\) and Eq.~\eqref{eq:spectrum}.

From \(H_{\OBC}=SH_eS^{-1}\), the canonical right eigenvector obeys
\begin{align}
H_{\OBC}S\ket{\phi_\mu}
&=SH_e\ket{\phi_\mu}
=2g\mu S\ket{\phi_\mu},
\label{eq:right_proof}
\end{align}
while \(H_{\OBC}^\dagger=S^{-1}H_eS\) gives
\begin{align}
H_{\OBC}^\dagger S^{-1}\ket{\phi_\mu}
&=S^{-1}H_e\ket{\phi_\mu}
=2g\mu S^{-1}\ket{\phi_\mu}.
\label{eq:left_proof}
\end{align}
Their overlap is
\begin{equation}
\langle\psi_\nu^L|\psi_\mu^R\rangle
=\langle\phi_\nu|S^{-1}S|\phi_\mu\rangle
=\delta_{\nu\mu}.
\label{eq:biorthonormality}
\end{equation}

For any nonzero complex \(c_\mu\), the transformation
\begin{equation}
\ket{\psi_\mu^R}\longrightarrow
c_\mu\ket{\psi_\mu^R},\qquad
\ket{\psi_\mu^L}\longrightarrow
(c_\mu^{-1})^*\ket{\psi_\mu^L}
\label{eq:mode_rescaling}
\end{equation}
preserves biorthonormality. Equation~\eqref{eq:RL} fixes this freedom by taking \(\ket{\phi_\mu}\) Euclidean normalized and \(s_1=1\). In this canonical convention, Hermitian completeness,
\begin{equation}
\sum_\mu\phi_\mu(j)\phi_\mu(\ell)^*=\delta_{j\ell},
\label{eq:completeness}
\end{equation}
immediately yields Eq.~\eqref{eq:integratedweights}. Under Eq.~\eqref{eq:mode_rescaling}, \(\mathcal W_R^{\rm can}\) and \(\mathcal W_L^{\rm can}\) change unless \(|c_\mu|=1\), whereas \(\mathcal W_{\rm bio}\), the resolvent, and the transfer amplitudes are unchanged.

Substituting Eq.~\eqref{eq:RL} into the rescaling-invariant definition of \(K_\mu\) gives Eq.~\eqref{eq:Petermann}.

Finally, \(S=S^\dagger\) and \(H_e=H_e^\dagger\) imply
\begin{equation}
H_{\OBC}^\dagger
=(S^{-1})^\dagger H_e^\dagger S^\dagger
=S^{-1}H_eS.
\label{eq:Hdagger}
\end{equation}
Using \(\eta=S^{-2}\),
\begin{align}
H_{\OBC}^\dagger\eta
&=S^{-1}H_eSS^{-2}=S^{-1}H_eS^{-1},\nonumber\\
\eta H_{\OBC}
&=S^{-2}SH_eS^{-1}=S^{-1}H_eS^{-1},
\label{eq:pseudo_steps}
\end{align}
which proves Eq.~\eqref{eq:pseudo}. Differentiating the metric norm of the propagator gives
\begin{align}
\frac{\mathrm d}{\mathrm dt}
\left(U^\dagger\eta U\right)
&=\ii U^\dagger H_{\OBC}^\dagger\eta U
-\ii U^\dagger\eta H_{\OBC}U=0.
\label{eq:pseudounitary_derivative}
\end{align}
Since \(U(0)=I\), Eq.~\eqref{eq:pseudounitary} follows.

\section{Binomial asymptotics and bounded-coupling scaling}
\label{app:scaling}

The binomial theorem gives the exact normalization
\begin{equation}
\sum_{j=1}^{N}s_j^2
=\sum_{k=0}^{M}\binom{M}{k}r^k=(1+r)^M.
\label{eq:binomial_normalization}
\end{equation}
With \(k=j-1\), division of Eq.~\eqref{eq:sprofile} by Eq.~\eqref{eq:binomial_normalization} gives
\begin{align}
\varpi_j
&=\frac{r^k\binom{M}{k}}{(1+r)^M}\nonumber\\
&=\binom{M}{k}
\left(\frac{r}{1+r}\right)^k
\left(\frac{1}{1+r}\right)^{M-k}\nonumber\\
&=\binom{M}{k}p^k(1-p)^{M-k}.
\label{eq:binomial_steps}
\end{align}
The standard binomial moments are \(\langle k\rangle=Mp\) and \(\operatorname{Var}(k)=Mp(1-p)\), which yield Eq.~\eqref{eq:centerwidth}. The gauge-field zero and the continuous envelope center differ by
\begin{equation}
j_c-j_\gamma=1-p.
\label{eq:center_difference}
\end{equation}
The discrete maximizer \(j_{\max}\) is an integer mode of the binomial law and need not equal either continuous coordinate.

For fixed \(p\in(0,1)\) and \(M\gg1\), the central-limit approximation is
\begin{equation}
\varpi_j\simeq
\frac{1}{\sqrt{2\pi\sigma_j^2}}
\exp\left[-\frac{(j-j_c)^2}{2\sigma_j^2}\right].
\label{eq:Gaussian}
\end{equation}
Approximating the sum of the squared envelope by an integral gives
\begin{align}
\mathcal I_S
&\simeq\int_{-\infty}^{\infty}
\frac{\mathrm dx}{2\pi\sigma_j^2}
\exp\left[-\frac{(x-j_c)^2}{\sigma_j^2}\right]\nonumber\\
&=\frac{1}{2\pi\sigma_j^2}\sqrt{\pi\sigma_j^2}
=\frac{1}{2\sqrt{\pi\sigma_j^2}}\nonumber\\
&=\frac{1}{2\sqrt{\pi(N-1)p(1-p)}}.
\label{eq:IPR_derivation}
\end{align}
Equations~\eqref{eq:Gaussian} and \eqref{eq:IPR_derivation} assume fixed \(p\in(0,1)\) separated from 0 and 1 as \(N\) grows. For a finite chain, an interior maximum additionally requires the strict interval in Eq.~\eqref{eq:focusing_regime}. The one-way limits lie outside both this interval and the central-limit approximation.

For positions away from the center, set \(x=(j-1)/M\). Stirling's formula applied to Eq.~\eqref{eq:binomial_steps} gives
\begin{equation}
\varpi_j\simeq
\frac{\exp[-M D(x\|p)]}{\sqrt{2\pi Mx(1-x)}},
\label{eq:large_deviation}
\end{equation}
where
\begin{equation}
D(x\|p)=x\ln\frac{x}{p}+(1-x)\ln\frac{1-x}{1-p}.
\label{eq:relative_entropy}
\end{equation}
This large-deviation form applies for \(0<x,p<1\), away from the interval endpoints. Expanding \(D(x\|p)\) to second order around \(x=p\) recovers Eq.~\eqref{eq:Gaussian}.

The maximum transformed hopping is
\begin{equation}
\tau_{\max}=g\max_{1\le j\le N-1}\sqrt{j(N-j)}
\simeq\frac{gN}{2}.
\label{eq:tau_max}
\end{equation}
To keep this coupling bounded, take
\begin{equation}
g=\frac{g_0}{N},\qquad g_0>0.
\label{eq:bounded_g}
\end{equation}
Then
\begin{equation}
\tau_j=\frac{g_0}{N}\sqrt{j(N-j)},
\qquad \tau_{\max}\longrightarrow\frac{g_0}{2},
\label{eq:bounded_tau}
\end{equation}
while
\begin{equation}
t_0=\frac{\pi}{2g}=\frac{\pi N}{2g_0}.
\label{eq:bounded_t0}
\end{equation}
The ratio \(r\), and hence \(\varpi_j\), is unchanged. The bounded-coupling family therefore preserves the spatial focus at the cost of a mirror time linear in \(N\).

\section{Propagator and resolvent derivations}
\label{app:propagator_resolvent}

\subsection{Full \texorpdfstring{Wigner-\(d\)}{Wigner-d} propagator}
\label{app:wigner}

For a real rotation angle \(\beta\), the spin operators satisfy
\begin{equation}
J_x=\ee^{\ii(\pi/2)J_z}J_y\ee^{-\ii(\pi/2)J_z}.
\label{eq:Jx_rotation}
\end{equation}
Consequently,
\begin{align}
\langle J,m'|\ee^{-\ii\beta J_x}|J,m\rangle
&=\ee^{\ii m'\pi/2}
\langle J,m'|\ee^{-\ii\beta J_y}|J,m\rangle
\ee^{-\ii m\pi/2}\nonumber\\
&=\ii^{m'-m}d^J_{m'm}(\beta),
\label{eq:x_rotation_matrix}
\end{align}
which fixes the phase convention in Eq.~\eqref{eq:wignerd}. The finite factorial sum used in the numerical scripts is
\begin{align}
d^J_{m'm}(\beta)
&=\sqrt{(J+m)!(J-m)!(J+m')!(J-m')!}\nonumber\\
&\quad\times\sum_k
\frac{(-1)^{m'-m+k}
[\cos(\beta/2)]^{2J+m-m'-2k}}
{(J+m-k)!k!}\nonumber\\
&\quad\times
\frac{[\sin(\beta/2)]^{m'-m+2k}}
{(m'-m+k)!(J-m'-k)!}.
\label{eq:wigner_sum}
\end{align}
The sum contains only integer \(k\) for which every factorial argument is nonnegative. Since \(J\pm m\) and \(J\pm m'\) are nonnegative integers, all factorials are ordinary integer factorials even when \(J\) is half integer. The scripts evaluate factorial ratios through \(\mathrm{gammaln}\) to reduce overflow.

At \(\beta=\pi\), the rotation satisfies
\begin{equation}
\ee^{-\ii\pi J_x}\ket{J,m}=(-\ii)^{2J}\ket{J,-m}.
\label{eq:mirror_rotation}
\end{equation}
Since \(m_{\bar j}=-m_j\) and \(2J=N-1\),
\begin{equation}
A_{\bar j,j}(t_0)=(-\ii)^{N-1}\frac{s_{\bar j}}{s_j}.
\label{eq:mirror_amplitude_appendix}
\end{equation}
The binomial symmetry
\begin{equation}
\binom{N-1}{\bar j-1}=\binom{N-1}{j-1}
\label{eq:binomial_mirror}
\end{equation}
then gives
\begin{equation}
\left(\frac{s_{\bar j}}{s_j}\right)^2
=r^{\bar j-j}=r^{N+1-2j},
\label{eq:mirror_gain_appendix}
\end{equation}
which proves Eqs.~\eqref{eq:mirrorgain} and \eqref{eq:gainproduct}.

Direct evaluation of Eq.~\eqref{eq:wigner_sum} agrees with matrix exponentiation of \(H_{\OBC}\) to numerical precision for the parameters used in Fig.~\ref{fig:mirror}.

\subsection{Green-function pole and residue derivation}
\label{app:green}

For \(z\) outside the spectrum,
\begin{align}
G(z)&=(zI-H_{\OBC})^{-1}\nonumber\\
&=[S(zI-H_e)S^{-1}]^{-1}\nonumber\\
&=S(zI-H_e)^{-1}S^{-1}.
\label{eq:green_similarity_steps}
\end{align}
Resolving the Hermitian partner in its orthonormal eigenbasis gives
\begin{equation}
G_{\ell j}(z)=\frac{s_\ell}{s_j}
\sum_{\mu=-J}^{J}
\frac{\phi_\mu(\ell)\phi_\mu(j)^*}{z-2g\mu}.
\label{eq:green_spectral}
\end{equation}
The coefficient of \((z-2g\mu)^{-1}\) is Eq.~\eqref{eq:residue}. At the endpoints, the Hermitian eigenvector weights can be chosen real and satisfy
\begin{equation}
|\phi_\mu(1)|^2=|\phi_\mu(N)|^2
=2^{-(N-1)}\binom{N-1}{J+\mu},
\label{eq:endpoint_weights}
\end{equation}
together with the mirror parity
\begin{equation}
\phi_\mu(N)=(-1)^{J-\mu}\phi_\mu(1).
\label{eq:endpoint_parity}
\end{equation}

To derive the endpoint Green function without summing Eq.~\eqref{eq:green_spectral}, apply Cramer's rule to the tridiagonal matrix \(zI-H_e\). Removing row 1 and column \(N\) leaves a triangular minor whose off-diagonal product is \((-1)^{N-1}\prod_{j=1}^{N-1}\tau_j\). The cofactor sign is \((-1)^{1+N}\), and the two signs cancel. Hence
\begin{equation}
G^{(e)}_{N1}(z)=
\frac{\displaystyle\prod_{j=1}^{N-1}\tau_j}
{\det(zI-H_e)}.
\label{eq:endpoint_cofactor}
\end{equation}
The hopping product is
\begin{align}
\prod_{j=1}^{N-1}\tau_j
&=g^{N-1}
\sqrt{\prod_{j=1}^{N-1}j
\prod_{j=1}^{N-1}(N-j)}\nonumber\\
&=g^{N-1}\sqrt{[(N-1)!]^2}
=g^{N-1}(N-1)!,
\label{eq:hopping_product}
\end{align}
and the determinant is
\begin{equation}
\det(zI-H_e)=\prod_{\mu=-J}^{J}(z-2g\mu).
\label{eq:green_determinant}
\end{equation}
Equations~\eqref{eq:endpoint_cofactor}-\eqref{eq:green_determinant} prove Eq.~\eqref{eq:endpointGreen}.

Since \((zI-H_e)^{-1}\) is symmetric,
\begin{align}
\frac{G_{\ell j}(z)}{G_{j\ell}(z)}
&=\frac{(s_\ell/s_j)G^{(e)}_{\ell j}(z)}
{(s_j/s_\ell)G^{(e)}_{j\ell}(z)}
=\left(\frac{s_\ell}{s_j}\right)^2.
\label{eq:green_ratio_steps}
\end{align}
For a mirror pair,
\begin{equation}
\frac{G_{\bar j,j}(z)}{G_{j,\bar j}(z)}
=r^{N+1-2j},
\qquad
C_G=2(N+1-2j)\ln r.
\label{eq:mirror_green_contrast}
\end{equation}
The same amplitude ratio holds for the absolute residues:
\begin{equation}
\frac{|R_{\bar j,j}^{(\mu)}|}{|R_{j,\bar j}^{(\mu)}|}
=r^{N+1-2j}.
\label{eq:mirror_residue_ratio}
\end{equation}
The ratios in Eqs.~\eqref{eq:green_ratio_steps}-\eqref{eq:mirror_residue_ratio} are defined when the denominator is nonzero. If the common Hermitian matrix element has an isolated zero, both directions vanish at that point and the ratio is understood through the surrounding analytic function. The broadening \(\kappa>0\) in \(z=E+\ii\kappa\) places the evaluation point in the upper half-plane and regularizes the finite-system poles.

\section{Disorder perturbation theory and numerical methods}
\label{app:disorder}

\begin{figure*}[!t]
\includegraphics[width=0.96\textwidth]{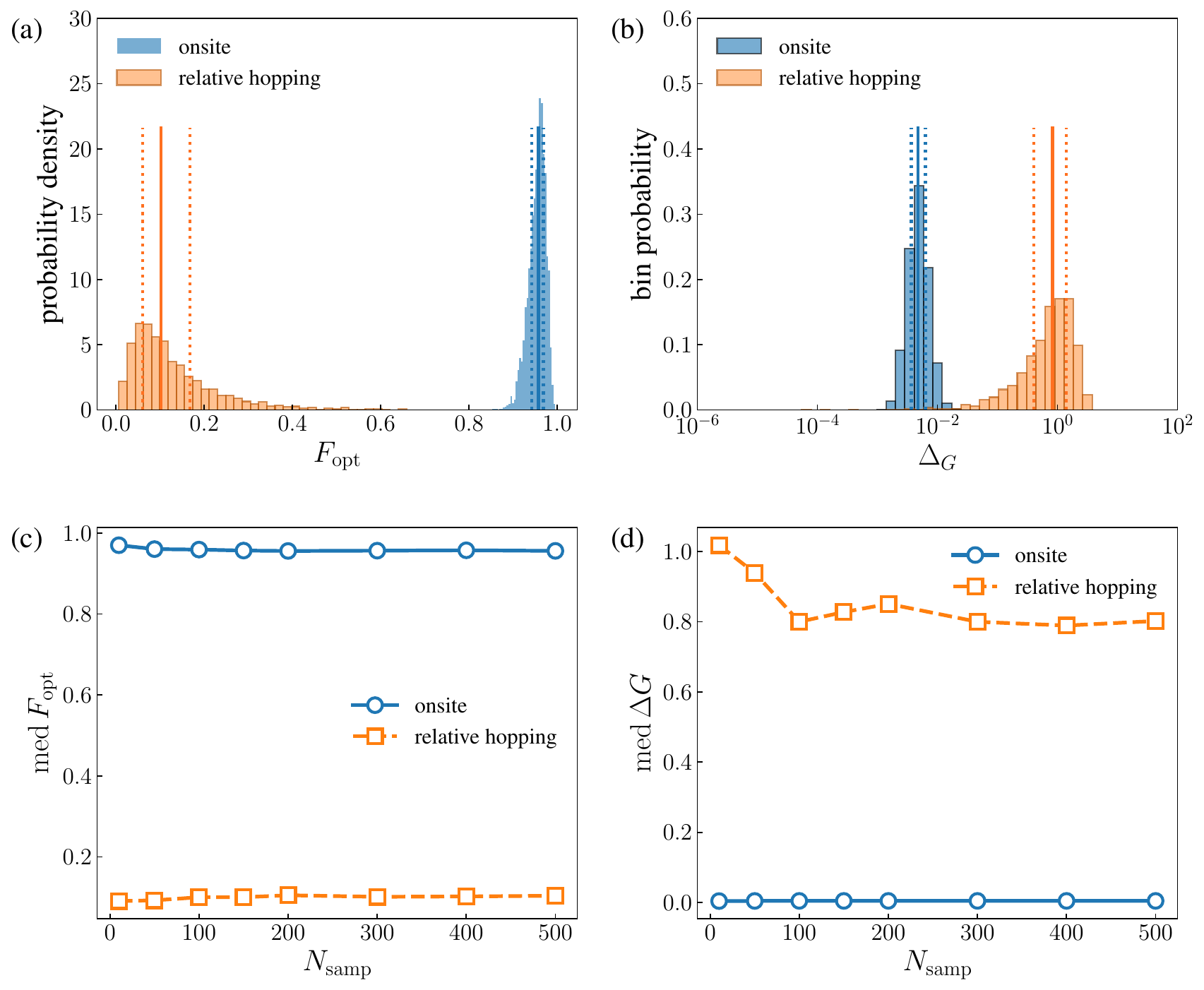}
\caption{Disorder distributions and sample convergence for \(N=24\), \(r=1.2\), \(W=0.30\), and seed 2701. (a) Probability densities of \(F_{\rm opt}\) and (b) logarithmically binned probabilities of \(\Delta_G\) for onsite and relative hopping disorder. The color and line-style key in panel (a) applies to all panels. Transparent fills distinguish the two ensembles. Solid vertical lines mark medians, and dotted vertical lines mark the first and third quartiles. Panels (c) and (d) show \(\operatorname{med}F_{\rm opt}\) and \(\operatorname{med}\Delta_G\), respectively, where the running median at sample count \(n\) is evaluated over the first \(n\) members of the fixed random sequence. The definitions of \(t_*\), \(F_{\rm opt}\), and \(G_{\rm opt}\) are identical to those used in Fig.~\ref{fig:disorder}.}
\label{fig:appendix_disorder}
\end{figure*}

Figure~\ref{fig:appendix_disorder} gives the full finite-disorder distributions and sample-number convergence for the diagnostics defined in Sec.~\ref{sec:disorder}; the perturbative and numerical definitions are detailed below.

For onsite disorder, first-order biorthogonal perturbation theory gives
\begin{align}
\delta E_\mu^{(\epsilon)}
&=\langle\psi_\mu^L|V_\epsilon|\psi_\mu^R\rangle\nonumber\\
&=\sum_j s_j^{-1}\phi_\mu(j)^*\epsilon_j
s_j\phi_\mu(j)\nonumber\\
&=\sum_j\epsilon_j|\phi_\mu(j)|^2.
\label{eq:onsite_shift}
\end{align}
The independent uniform variables have
\begin{equation}
\langle\epsilon_j\rangle=0,\qquad
\operatorname{Var}(\epsilon_j)=\frac{W_\epsilon^2g^2}{3}.
\label{eq:onsite_statistics}
\end{equation}
Independence removes every cross term, so
\begin{align}
\operatorname{Var}(\delta E_\mu^{(\epsilon)})
&=\sum_j\operatorname{Var}(\epsilon_j)|\phi_\mu(j)|^4\nonumber\\
&=\frac{W_\epsilon^2g^2}{3}\sum_j|\phi_\mu(j)|^4.
\label{eq:onsite_variance}
\end{align}

The hopping perturbation is
\begin{equation}
\delta H_h=\sum_{j=1}^{N-1}\left[
u_j\xi_j^u\ket j\bra{j+1}
+v_j\xi_j^v\ket{j+1}\bra j\right].
\label{eq:hopping_perturbation}
\end{equation}
Choosing the real eigenvectors of \(H_e\), its first-order shift is
\begin{align}
\delta E_\mu^{(h)}
&=\sum_j\left[
u_j\xi_j^u\frac{s_{j+1}}{s_j}
+v_j\xi_j^v\frac{s_j}{s_{j+1}}\right]
\phi_\mu(j)\phi_\mu(j+1)\nonumber\\
&=\sum_j\tau_j(\xi_j^u+\xi_j^v)
\phi_\mu(j)\phi_\mu(j+1).
\label{eq:hopping_shift}
\end{align}
Since \(\operatorname{Var}(\xi_j^u+\xi_j^v)=2W_h^2/3\),
\begin{equation}
\operatorname{Var}(\delta E_\mu^{(h)})
=\frac{2W_h^2}{3}\sum_{j=1}^{N-1}
\tau_j^2|\phi_\mu(j)\phi_\mu(j+1)|^2.
\label{eq:hopping_variance}
\end{equation}

For any realization with positive disordered bonds \(\widetilde u_j=u_j(1+\xi_j^u)\) and \(\widetilde v_j=v_j(1+\xi_j^v)\), an OBC symmetrizing map still exists and obeys
\begin{equation}
\frac{\widetilde s_{j+1}}{\widetilde s_j}
=\sqrt{\frac{\widetilde v_j}{\widetilde u_j}}.
\label{eq:disordered_similarity}
\end{equation}
It is sample dependent, and the symmetric hopping becomes
\begin{equation}
\sqrt{\widetilde u_j\widetilde v_j}
=\tau_j\sqrt{(1+\xi_j^u)(1+\xi_j^v)},
\label{eq:disordered_symmetric_hopping}
\end{equation}
which is not generally proportional to \(\sqrt{j(N-j)}\) with a bond-independent coefficient. Relative hopping disorder therefore breaks the clean analytic Krawtchouk map and its exactly commensurate mirror chain, rather than diagonal symmetrizability itself.

The finite-disorder calculations use \(N=24\), \(\alpha_1=1\), \(r=1.2\), \(j_0=5\), \(\kappa=0.12g\), and the fixed Mersenne-Twister seed 2701. Eleven disorder strengths span \(0.005\le W\le1\) on a logarithmic grid. At each strength and for each disorder type, 500 independent realizations are generated. Time propagation uses direct matrix exponentiation on an 81-point grid over \(0.8t_0\le t\le1.2t_0\). Resolvents are evaluated by linear solves rather than explicit matrix inversion.

For the ladder diagnostic, let \(E_n^{\rm dis}\) be the disordered eigenvalues and \(E_\mu^{(0)}=2g\mu\) the clean ladder. The assignment \(\pi_*\) minimizes
\begin{equation}
\pi_*=\underset{\pi}{\arg\min}
\sum_\mu|E_{\pi(\mu)}^{\rm dis}-E_\mu^{(0)}|^2.
\label{eq:optimal_assignment}
\end{equation}
The minimum-cost assignment is obtained with a self-contained Hungarian algorithm. Complex coefficients \(\beta_0\) and \(\beta_1\) are then found by least squares, and
\begin{equation}
\Delta_{\rm ladder}=
\frac{1}{2g}
\left[\frac1N\sum_\mu
|E_{\pi_*(\mu)}^{\rm dis}-(\beta_0+\beta_1\mu)|^2\right]^{1/2}.
\label{eq:ladder_error}
\end{equation}
For weak and moderate disorder, this is a matched distortion of the clean commensurate ladder. At strong disorder it is retained only as an empirical spectral measure.

The plotted central curves are sample medians, and the shaded bands are the 25th to 75th percentile intervals. Sample-number convergence compares the first 200 members of the same random sequence with all 500 members, isolating convergence from changes in the ensemble or seed. Figure~\ref{fig:appendix_disorder} shows the full distributions at \(W=0.30\) and the convergence of the running medians.

\FloatBarrier

\section{Loop obstruction and high-order exceptional point}
\label{app:loop_ep}

Under the open-chain map, the two closing terms in Eq.~\eqref{eq:Hb} become
\begin{align}
S^{-1}\left(w_{1\leftarrow N}\ket1\bra N\right)S
&=w_{1\leftarrow N}\frac{s_N}{s_1}\ket1\bra N,
\label{eq:closure_forward}\\
S^{-1}\left(w_{N\leftarrow1}\ket N\bra1\right)S
&=w_{N\leftarrow1}\frac{s_1}{s_N}\ket N\bra1.
\label{eq:closure_reverse}
\end{align}
The ratio of the transformed amplitudes is
\begin{align}
\frac{\widetilde w_{1\leftarrow N}}
{\widetilde w_{N\leftarrow1}}
&=\frac{w_{1\leftarrow N}}{w_{N\leftarrow1}}
\frac{s_N^2}{s_1^2}\nonumber\\
&=\frac{w_{1\leftarrow N}}{w_{N\leftarrow1}}
\prod_{j=1}^{N-1}\frac{v_j}{u_j}\nonumber\\
&=\exp\left(2\sum_{j=1}^{N-1}\gamma_j
+\ln\frac{w_{1\leftarrow N}}{w_{N\leftarrow1}}\right)\nonumber\\
&=\ee^{2\Phi_{\rm im}}.
\label{eq:closure_ratio}
\end{align}
The product of the two amplitudes is invariant and equals \(\tau_b^2\). Therefore
\begin{equation}
\widetilde w_{1\leftarrow N}=\tau_b\ee^{\Phi_{\rm im}},
\qquad
\widetilde w_{N\leftarrow1}=\tau_b\ee^{-\Phi_{\rm im}}.
\label{eq:closure_amplitudes}
\end{equation}
Hermiticity requires these positive amplitudes to be equal, which is equivalent to \(\Phi_{\rm im}=0\). This is the loop-consistency condition that is absent under OBC.

The eigenvalue trajectories in Fig.~\ref{fig:loop_flux}(a) are evaluated in the transformed representation \(H_e+H_b(\Phi_{\rm im})\), which is isospectral to the physical ring. At each adjacent flux step, the two spectra are matched by minimizing the total squared complex-plane displacement with the Hungarian algorithm. Repeating the calculation with half the flux step tests the geometric trajectory set. Branch labels can permute at exact crossings, but no independently sorted eigenvalue lists are connected.

For the upper-triangular one-way limit, repeated multiplication of
\begin{equation}
H_{\OBC}=\alpha_1\sum_{j=1}^{N-1}j\ket j\bra{j+1}
\label{eq:one_way_H}
\end{equation}
gives
\begin{equation}
\begin{aligned}
H_{\OBC}^q={}&\alpha_1^q
\sum_{j=1}^{N-q}\frac{(j+q-1)!}{(j-1)!}
\ket j\bra{j+q},\\
&q=1,\ldots,N-1.
\end{aligned}
\label{eq:H_power}
\end{equation}
To verify Eq.~\eqref{eq:H_power}, assume it holds at power \(q\). Left multiplication by Eq.~\eqref{eq:one_way_H} extends each surviving path by one bond and multiplies its coefficient by the next integer, producing \((j+q)!/(j-1)!\) at power \(q+1\). At \(q=N-1\), only \(j=1\) remains, giving the nonzero result in Eq.~\eqref{eq:nilpotent}; at \(q=N\), no path remains. Moreover, \(\operatorname{rank}(H_{\OBC}^q)=N-q\) and \(\dim\ker(H_{\OBC}^q)=q\). These kernel dimensions are those of one nilpotent Jordan block of size \(N\), which proves the order-\(N\) exceptional point.

The conditioning curve is computed from
\begin{equation}
\log_{10}\kappa_S=
\frac{\max_jX_j-\min_jX_j}{\ln10},
\label{eq:log_condition}
\end{equation}
so no exponentially large number is formed in double precision. For \(r<1/(N-1)\), \(s_1\) is the largest entry and \(s_N=r^{(N-1)/2}\) is the smallest, giving the exact small-\(r\) form
\begin{equation}
\log_{10}\kappa_S
=-\frac{N-1}{2}\log_{10}r.
\label{eq:condition_asymptotic}
\end{equation}
At \(N=12\) and \(r=10^{-8}\), this yields \(\log_{10}\kappa_S=44\). The validation script independently evaluates the diagonal entries with 80-digit arithmetic when the Symbolic Math Toolbox is available and otherwise checks Eq.~\eqref{eq:log_condition} against the exact small-\(r\) expression. Thus no ordinary double-precision eigenvector condition number is used as quantitative evidence near the exceptional point.

\begin{figure*}[!t]
\includegraphics[width=0.96\textwidth]{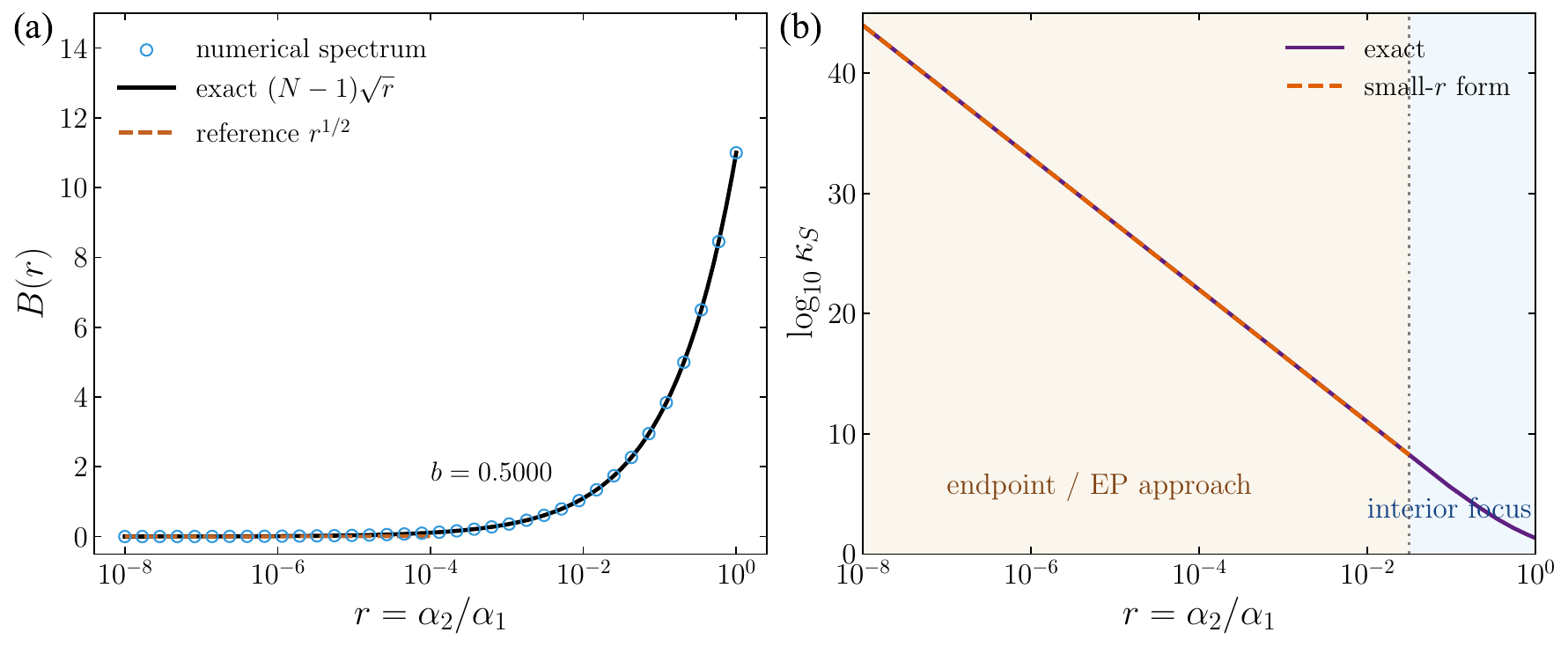}
\caption{Approach to the one-way exceptional point for \(N=12\). (a) Spectral half-width \(B(r)=\alpha_1^{-1}\max_\mu|E_\mu|\). Numerical eigenvalues coincide with \((N-1)\sqrt r\), with fitted logarithmic slope \(0.5000\). (b) Exact canonical \(\log_{10}\kappa_S\). The dashed small-\(r\) form is restricted to \(r<r_c/3\); the vertical line at \(r_c=1/(N-1)\) and shading distinguish interior from endpoint focusing. The exceptional point at \(r=0\) is omitted from the logarithmic axis.}
\label{fig:appendix_ep}
\end{figure*}

For fixed \(\alpha_1\) and \(r>0\), the spectral half-width is
\begin{equation}
B(r)\equiv\frac{1}{\alpha_1}\max_\mu|E_\mu|
=(N-1)\sqrt r.
\label{eq:spectral_half_width}
\end{equation}
Figure~\ref{fig:appendix_ep} compares this exact spectral collapse with the log-space conditioning in Eqs.~\eqref{eq:log_condition} and \eqref{eq:condition_asymptotic}. The crossover \(r_c=1/(N-1)\) marks the finite-chain boundary between an interior maximum of the similarity envelope and its left-endpoint maximum along the approach to \(r=0\). The point \(r=0\), which cannot be included on the logarithmic axis, is the exact order-\(N\) exceptional point.

\FloatBarrier

\balance
%\bibliography{references}

\end{document}